\documentclass[english]{article}
\usepackage[T1]{fontenc}
\usepackage[latin9]{inputenc}
\usepackage{babel}
\usepackage{amsmath}
\usepackage{amssymb}
\usepackage{esint}
\usepackage[unicode=true,pdfusetitle,
 bookmarks=true,bookmarksnumbered=false,bookmarksopen=false,
 breaklinks=false,pdfborder={0 0 0},backref=false,colorlinks=false]
 {hyperref}
\usepackage{breakurl}

\makeatletter
\usepackage{babel}

\usepackage{color}
\usepackage{babel}
\usepackage{breakurl}
\usepackage{babel}
\usepackage{babel}
\usepackage{babel}
\usepackage{amsfonts}
\usepackage{babel}
\usepackage{babel}
\usepackage{graphicx}

\makeatother

\begin{document}

\title{Maxwell field with Torsion}

\author{R. Fresneda%
\thanks{email:rodrigo.fresneda@ufabc.edu.br%
}\medskip{}
 \\
{\small{}Federal University of ABC (UFABC)}\\
{\small{} Av. dos Estados, 5001. Bairro Bangu. }\\
{\small{}Santo André - SP - Brasil . CEP 09210-580,} \medskip{}
 \\
 M.C. Baldiotti%
\thanks{baldiotti@uel.br%
}, T.S. Pereira%
\thanks{tspereira@uel.br%
}\medskip{}
 \\
{\small{}State University of Londrina (UEL)}\\
{\small{} Departamento de Física}\\
 {\small{}Rodovia Celso Garcia Cid (Pr 445), }\\
 {\small{}Km 380 CEP 86057-970, }\\
 {\small{}Caixa Postal 6001 Londrina-PR, Brazil}\\
 }
\maketitle
\begin{abstract}
We propose a generalizing gauge-invariant model of propagating torsion
which couples to the Maxwell field and to charged particles. As a
result we have an Abelian gauge invariant action which leads to a
theory with nonzero torsion and which is consistent with available
experimental data. 
\end{abstract}

\section{Introduction}

In gravitation theories the Minimal Coupling Procedure (MCP) can be
simply stated as a procedure which, starting from a theory in flat
spacetime, substitutes all partial derivatives by covariant derivatives
and the flat metric by the Riemannian metric. The impossibility of
achieving simultaneously the usual gauge invariance of electromagnetism
and MCP of the gauge field to torsion has led many authors to abandon
the MCP, keeping usual partial derivatives (as opposed to the covariant
ones) in the definition of the electromagnetic field tensor \cite{Shapiro2002}.
However, perturbative results from QFT of the photon self-energy suggest
corrections to the Maxwell equations coming from the coupling between
spinors and an external torsion field \cite{DeSabbata1980}. Together
with the fact that in Riemann-Cartan spacetime the particle spin also
works as a source for torsion \cite{Hehl1976}, one is tempted to
modify the electromagnetic field tensor in a minimal way.

In \cite{Novello1976} only the trace part of the torsion tensor minimally
couples to the electromagnetic field tensor. Charge conservation and
gauge invariance constrain the torsion trace to be a gradient of a
constant scalar field, thus leading to a trivial theory with non-propagating
torsion. Among the ``minimal\textquotedblright{} non-trivial modifications,
we single out two formalisms. The first one amounts to a modification
of the gauge transformation, keeping the usual MCP \cite{Hojman1978,Hojman1979}:
\begin{equation}
\delta_{\epsilon}A_{\mu}=e^{\phi}\partial_{\mu}\epsilon\,,\label{eq:HRRS-gauge}
\end{equation}
where $\phi\left(x\right)$ is a scalar field whose gradient gives
the trace part of the torsion tensor. This formalism is known
as the \emph{Hojman-Rosenbaum-Ryan-Shepley propagating torsion} (HRRS
model) and has been developed, in particular, for non-Abelian gauge
theories in \cite{Mukku1979}.

The second formalism, which we refer to as Saa's theory, represents
a change in the MCP procedure by modification of the definition of
the invariant volume form \cite{Saa1993,Saa1997,Saa:dilaton}. This formalism,
which has been adopted by some authors \cite{Popawski2006}, does
not apply MCP for the electromagnetic field tensor, keeping the usual
definition in terms of the exterior derivative of the Maxwell gauge
connection. As a result, it does not violate usual $U\left(1\right)$
gauge symmetry. Later in \cite{Mosna2005}, by requiring that equivalence
classes of Lagrangians be related by MCP, it was shown that the trace
of the torsion tensor must be a gradient whose scalar deforms the
invariant volume form, exactly as was proposed before on geometrical
grounds.

Both formalisms provide a wave equation for a scalar whose gradient
is the trace part of the torsion tensor and which additionally involves
the electromagnetic field. These models, where the torsion's dynamics
is given by a wave equation, are called propagating torsion models.
In \cite{Popawski2006} a conformal transformation is made in order
to remove the negative sign of the kinetic term arising from the torsionic
scalar in Saa's theory. As a result, the invariant volume simply becomes
the square root of the determinant of the conformal metric, and MCP
is performed to the Maxwell field tensor in the sense of HRRS theory.
Despite their elegance, both formalisms are incompatible with experimental
results. In the case of HRRS theory, experimental data from the solar
system invalidate any predicted geodesic deviation of atoms with different
electromagnetic content \cite{Ni1979}. Fiziev et al. point out in
\cite{Boyadjiev1999,Fiziev1998b} that Saa's theory violates basic
experimental gravitational and solar system data.

The absence of any free parameters both in Saa's theory and in the
HRRS model, together with the experimental data, invalidates both
models. In this work we couple electromagnetism to torsion by a generalizing
Abelian gauge invariance to obtain MCP at the Action level. Here we
propose an alternate model to HRRS's, where compatibility of a new
gauge principle with MCP requires the introduction of a free parameter,
which is constrained by the available experimental data. As a result
we have an Abelian gauge invariant Action which leads to a theory
of propagating torsion with nonzero trace part and which is consistent
with available experimental data. Besides, the theory we present also
admits a formulation in terms of Semi-Minimal Coupling, i.e., MCP
at the level of differential forms. Therefore, our proposal effectively
rehabilitates a model of propagating torsion closely related to previous
attempts.

The work is organized as follows: in Section 2 we introduce notation
and definitions used throughout this work and also describe in detail
the compatibility between gauge invariance and MCP. In Section 3 we
present a new gauge transformation which provides an invariant Maxwell-like
action and we propose a redefinition of the trace part of the torsion
tensor in order to achieve MCP at the action level. In this connection
we also discuss coupling to scalar fields and the Newtonian limit.
In Section 4 we conclude with some final remarks and perspectives.
In the Appendix we provide the semi-minimal coupling procedure formalism,
and we apply it to obtain sources to the Maxwell equations.

Throughout this work we use units in which $c=G=\hbar=1$ and metric
signature $\left(+,-,-,-\right)$.

\section{Gauge invariance and MCP compatibility}

We use the following definitions for torsion and covariant derivative.
Let $\nabla$ be an affine connection compatible with the Lorentzian
metric $g$. The covariant derivative of a vector field $\upsilon$
along $\partial_{\mu}$ in a local chart $x^{\mu}$ is 
\begin{equation}
\nabla_{\mu}\upsilon^{\nu}=\partial_{\mu}\upsilon^{\nu}+\Gamma_{\mu\sigma}^{\nu}\upsilon^{\sigma}.\label{eq:covariant-derivative}
\end{equation}
The tensor components of the torsion of the connection $\nabla$ can
be given in terms of connection coefficients as 
\[
T_{\mu\nu}{}^{\sigma}=\Gamma_{\mu\nu}{}^{\sigma}-\Gamma_{\nu\mu}{}^{\sigma}\,.
\]
The contorsion tensor is defined by 
\[
K_{\mu\nu\sigma}=\frac{1}{2}\left(T_{\mu\nu\sigma}+T_{\sigma\nu\mu}+T_{\sigma\mu\nu}\right)=-K_{\mu\sigma\nu}\Rightarrow T_{\mu\nu\sigma}=K_{\mu\nu\sigma}-K_{\nu\mu\sigma}\,.
\]
In terms of the contorsion tensor, one can split the covariant derivative
(\ref{eq:covariant-derivative}) in two parts, one involving the Levi-Civita
connection coefficients $\bar{\Gamma}$ of the Levi-Civita connection
$\bar{\nabla}$ and another involving the contorsion tensor, 
\begin{align}
\nabla_{\mu}\upsilon^{\nu} & =\partial_{\mu}\upsilon^{\nu}+\bar{\Gamma}_{\mu\sigma}{}^{\nu}\upsilon^{\sigma}+K_{\mu\sigma}{}^{\nu}\upsilon^{\sigma}\nonumber \\
 & =\bar{\nabla}_{\mu}\upsilon^{\nu}+K_{\mu\sigma}{}^{\nu}\upsilon^{\sigma}\,.\label{eq:covariant-derivative2}
\end{align}

We note that unlike in Saa's use of a transposed connection in \cite{Saa1997},
we adopt the usual definition for the covariant derivative of a scalar
density $S$ of weight $w$, i.e., 
\[
\nabla_{\mu}S=\partial_{\mu}S+w\Gamma_{\mu\sigma}{}^{\sigma}S=\partial_{\mu}S+w\bar{\Gamma}_{\mu\sigma}{}^{\sigma}S\,,
\]
since the contortion tensor is antisymmetric in the last two indices.
Therefore, we do not change the volume element, $\sqrt{-g}$ being
a density of weight $-1$, 
\[
\nabla_{\mu}\sqrt{\left\vert \det g\right\vert }=0\,.
\]

Quite generally, the torsion tensor can be divided in three parts,
\begin{equation}
T_{\mu\nu\sigma}=\frac{1}{3}\left(g_{\mu\sigma}T_{\nu}-g_{\nu\sigma}T_{\mu}\right)+\frac{1}{6}\varepsilon_{\mu\nu\sigma\rho}S^{\rho}+Q_{\mu\nu\sigma}\,,\label{eq:decomp-geral-torcao}
\end{equation}
such that $S_{\mu}$ is the axial part (also known as pseudo-trace),
$T_{\mu}$ is the trace part, and $Q_{\mu\nu\sigma}$ is tensor with
vanishing trace and pseudo-trace, i.e., $\varepsilon^{\mu\nu\sigma\kappa}Q_{\nu\sigma\kappa}=Q_{\sigma\mu}{}^{\sigma}=0$.
Therefore, the trace part $T_{\mu}$ is given by 
\begin{equation}
T_{\mu}=T_{\sigma\mu}{}^{\sigma}=K_{\sigma\mu}{}^{\sigma}\,,\label{eq:torsion-trace}
\end{equation}
while the axial part or pseudo-trace is given by 
\[
S^{\mu}=\varepsilon^{\mu\nu\sigma\kappa}T_{\nu\sigma\kappa}\,.
\]

Restricting, for now, the analysis to the homogeneous Maxwell equations
(without sources), the MCP leads to 
\[
\int d^{4}xf_{\mu\nu}f^{\mu\nu}\rightarrow\int d^{4}x\sqrt{-g}\tilde{F}_{\mu\nu}\tilde{F}^{\mu\nu},
\]
where $f_{\mu\nu}=\partial_{\mu}A_{\nu}-\partial_{\nu}A_{\mu}$ is
the usual electromagnetic field tensor and 
\begin{equation}
\tilde{F}_{\mu\nu}=\nabla_{\mu}A_{\nu}-\nabla_{\nu}A_{\mu}\,,\label{eq:Fmunu-a}
\end{equation}
is the one obtained via MCP. Using the definition of covariant derivative
in (\ref{eq:covariant-derivative2}) and the expression (\ref{eq:torsion-trace})
we have 
\begin{equation}
\tilde{F}_{\mu\nu}=f_{\mu\nu}-T_{\mu\nu}{}^{\sigma}A_{\sigma}\,.\label{eq:Fmunu}
\end{equation}
The expression for the minimally-coupled field tensor shows explicit
dependence on the potential, making MCP incompatible with usual $U\left(1\right)$
gauge invariance in the presence of torsion. In addition, in the case
where the torsion tensor is completely anti-symmetric, i.e., in the
presence of the torsion pseudo-trace alone, the homogeneous Maxwell
equation $\nabla_{\mu}\tilde{F}^{\mu\nu}=0$ reduces to 
\begin{align*}
\nabla_{\mu}\tilde{F}^{\mu\nu} & =\bar{\nabla}_{\mu}f^{\mu\nu}-\frac{1}{12}\varepsilon^{\mu\rho\nu\sigma}\left(\bar{\nabla}_{\mu}S_{\rho}-\bar{\nabla}_{\rho}S_{\mu}\right)A_{\sigma}-\frac{g}{36}\left(g^{\nu\sigma}S^{2}-S^{\sigma}S^{\nu}\right)A_{\sigma}+\\
 & +\frac{1}{6}\varepsilon^{\mu\kappa\nu\rho}f_{\mu\kappa}S_{\rho}=0\,.
\end{align*}
The non-invariant terms, linear in the gauge potential, 
\[
-\frac{1}{12}\varepsilon^{\mu\rho\nu\sigma}\left(\bar{\nabla}_{\mu}S_{\rho}-\bar{\nabla}_{\rho}S_{\mu}\right)A_{\sigma}-\frac{g}{36}\left(g^{\nu\sigma}S^{2}-S^{\sigma}S^{\nu}\right)A_{\sigma}\,,
\]
must vanish independently of $A_{\mu}$ and of each other. If one
contracts them with $A_{\nu}$, one finds a relation between torsion
and the gauge potential which is furthermore not invariant, namely,
\[
S^{2}A^{2}=\left(S_{\mu}A^{\mu}\right)^{2}\,.
\]
Therefore, the symmetric matrix $g^{\nu\sigma}S^{2}-S^{\sigma}S^{\nu}$
must vanish. Since it can be diagonalized to $\mathrm{diag}\left(0,S^{2},S^{2},S^{2}\right)$,
one arrives at the conclusion that $S^{2}=0$. Substituting this condition
back into the matrix, one is left with the trivial solution $S_{\mu}=0$.
As a result, it is not possible to perform MCP and maintain $U(1)$
gauge invariance for non vanishing torsion pseudo-trace.

Even in the case where the pseudo-trace vanishes, from (\ref{eq:Fmunu})
we thus come to the conclusion that is not possible to attain MCP,
unless one is willing to modify either the usual $U\left(1\right)$
gauge invariance of electromagnetism or the notion of MCP. In effect,
in \cite{Saa1993}, a different notion of MCP was used, where one
only demands MCP at the level of differential forms, with the exterior
derivative being replaced by a covariant exterior derivative ($d\rightarrow D$),
called ``Semi-Minimal Coupling Procedure'' (SMCP). While in \cite{Hojman1978},
MCP at the Action level was retained at the cost of modifying the
gauge invariance. In the Appendix \ref{sub:Semi-minimal-coupling}
we discuss a version of the SMCP applied to our theory.

\section{New gauge transformation and minimal coupling\label{sec:Minimal-coupling}}

Now consider the (global) gauge transformation with constant parameter
$\epsilon$ given by 
\begin{equation}
\delta_{\epsilon}A_{\mu}=T_{\mu}\epsilon\,,\label{eq:global-transf}
\end{equation}
where $T_{\mu}$ is in principle any four-vector. One can easily see
that in order that (\ref{eq:global-transf}) be a global symmetry
of the Maxwell action 
\[
\mathcal{A}_{0}=\frac{1}{4}\int d^{4}x\sqrt{-g}f^{2}\,,\,\, f_{\mu\nu}=\partial_{\mu}A_{\nu}-\partial_{\nu}A_{\mu}\,,
\]
it is necessary that $T_{\mu}$ be the gradient of a scalar function,
i.e., 
\begin{equation}
T_{\mu}=\partial_{\mu}\varphi\,,\label{eq:def-traco-torcao}
\end{equation}
since $\delta_{\epsilon}f_{\mu\nu}=\left(\partial_{\mu}T_{\nu}-\partial_{\nu}T_{\mu}\right)\epsilon$.
In order to obtain a local transformation, we promote $\epsilon$
to a function $\epsilon\left(x\right)$ of the spacetime coordinates.
Now the Maxwell action $A_{0}$ is no longer invariant, and the non
invariant term is 
\[
\delta_{\epsilon}\mathcal{A}_{0}=\int d^{4}x\sqrt{-g}f^{\mu\nu}\partial_{\nu}\varphi\partial_{\mu}\epsilon\,.
\]
The appearance of the derivatives $\partial_{\mu}\epsilon$ can be
compensated by a proper modification of the original gauge transformation
(\ref{eq:global-transf}). In what follows we use the well-known Noether
technique in constructing supergravity theories \cite{West1990},
in this case weaving together the usual $U\left(1\right)$ transformation
with the torsion field $T_{\mu}$. We then compensate non-invariant
terms order by order in a new parameter, in order to achieve an invariant
Lagrange functional. Now let 
\begin{equation}
\delta_{\epsilon}A_{\mu}=\frac{1}{\alpha}\partial_{\mu}\epsilon+\partial_{\mu}\varphi\epsilon\,,\label{eq:local-gauge-transf}
\end{equation}
be a local gauge transformation with parameter $\epsilon\left(x\right)$,
and some coupling constant $\alpha$. To order $\alpha^{0}$, the
action 
\[
\mathcal{A}_{1}=\mathcal{A}_{0}-\alpha\int d^{4}x\sqrt{-g}f^{\mu\nu}\partial_{\nu}\varphi A_{\mu}
\]
is invariant under (\ref{eq:local-gauge-transf}), and the non invariant
contribution is 
\[
\delta_{\epsilon}\mathcal{A}_{1}=-\alpha\int d^{4}x\sqrt{-g}\left(\partial^{\nu}\varphi\partial_{\nu}\varphi A_{\mu}-\partial^{\nu}\varphi A_{\nu}\partial_{\mu}\varphi\right)\partial^{\mu}\epsilon\,.
\]
We compensate this term with the addition of a new term to the action
$\mathcal{A}_{1}$, giving the total action 
\[
\mathcal{A}_{2}=\mathcal{A}_{1}+\frac{\alpha^{2}}{2}\int d^{4}x\sqrt{-g}\left(\partial^{\nu}\varphi\partial_{\nu}\varphi A^{\mu}A_{\mu}-\partial^{\mu}\varphi A_{\mu}\partial^{\nu}\varphi A_{\nu}\right)\,.
\]
One can now check that action $\mathcal{A}_{2}$ is invariant under
(\ref{eq:local-gauge-transf}) to all orders of $\alpha$, i.e., $\delta_{\epsilon}\mathcal{A}_{2}=0$.
Action $\mathcal{A}_{2}$ can be conveniently written as 
\begin{equation}
\mathcal{A}_{2}=\frac{1}{4}\int d^{4}x\sqrt{-g}F^{2}\,,\label{eq:A2}
\end{equation}
where 
\begin{equation}
F_{\mu\nu}=f_{\mu\nu}+\alpha\left(\partial_{\mu}\varphi A_{\nu}-\partial_{\nu}\varphi A_{\mu}\right)\,.\label{eq:field-tensor}
\end{equation}
Invariance of $\mathcal{A}_{2}$ can be easily seen from the identity
$\delta_{\epsilon}F_{\mu\nu}=0$. In addition, the equations of motion
that follow from action $\mathcal{A}_{2}$ are 
\begin{equation}
\left(\bar{\nabla}_{\mu}-\alpha T_{\mu}\right)F^{\mu\nu}=0\,,\label{eq:equations-motion}
\end{equation}
where $\bar{\nabla}$ is the Levi-Civita connection (no torsion).

The theory with action (\ref{eq:A2}) is gauge-invariant by the local
Abelian transformation (\ref{eq:local-gauge-transf}) and generally
covariant. However, at this point it is not clear that the vector
field $T_{\mu}$ plays the role of torsion. Below we provide the necessary
interpretation.

In order to give the interpretation that the model proposed in this
section actually describes a coupling between electromagnetism and
torsion, we turn to the definition of the Maxwell field tensor given
in terms of MCP (\ref{eq:Fmunu-a}), i.e., $\tilde{F}_{\mu\nu}=\nabla_{\mu}A_{\nu}-\nabla_{\nu}A_{\mu}$.
We wish to identify $\tilde{F}_{\mu\nu}$ with $F_{\mu\nu}$ from
(\ref{eq:field-tensor}). This can be achieved by rescaling the trace
components of the torsion tensor, such that now, instead of (\ref{eq:decomp-geral-torcao}),
one has for the trace part: 
\begin{equation}
T_{\mu\nu\sigma}=\alpha\left(g_{\mu\sigma}T_{\nu}-g_{\nu\sigma}T_{\mu}\right)\,.\label{eq:newdef}
\end{equation}
As a result, one has $\nabla_{\mu}A_{\nu}-\nabla_{\nu}A_{\mu}=F_{\mu\nu}$,
with $F_{\mu\nu}$ given by (\ref{eq:field-tensor}). Therefore, by
demanding MCP at the level of the Action, one is led to a nontrivial
coupling between torsion and electromagnetism, which manifests itself
in the very definition of the torsion tensor by means of the introduction
of the constant $\alpha$.

\subsection{Action functional with matter fields}

Given the local transformation (\ref{eq:local-gauge-transf}), let
us obtain a unitary realization of $U\left(1\right)$ on scalar fields
$\phi$, $\phi^{\prime}=\gamma\phi$, $\gamma^{\dagger}=\gamma^{-1}$,
such that the covariant derivative of the scalar field $\phi$ transforms
in the same representation, 
\[
D_{\mu}\phi\rightarrow\gamma D_{\mu}\phi\,.
\]

We search for a solution in the form 
\[
\gamma=e^{iqf\left(\varphi\right)\epsilon}\,,\,\, D_{\mu}\phi=\left(\partial_{\mu}-iqg\left(\varphi\right)A_{\mu}\right)\phi\,,
\]
where $f\left(\varphi\right)$ and $g\left(\varphi\right)$ are functions
of $\varphi$, $T_{\mu}=\partial_{\mu}\varphi$, $q$ is the charge
and $\epsilon$ is an arbitrary transformation parameter. Thus 
\begin{align*}
\left(D_{\mu}\phi\right)^{\prime} & =\partial_{\mu}\phi^{\prime}-iqg\left(\varphi\right)A_{\mu}^{\prime}\phi^{\prime}\\
 & =\gamma\partial_{\mu}\phi+iq\gamma\left(\epsilon\partial_{\mu}f+f\partial_{\mu}\epsilon\right)\phi-i\gamma qg\left(\varphi\right)\left(A_{\mu}+\frac{1}{\alpha}\partial_{\mu}\epsilon+\partial_{\mu}\varphi\epsilon\right)\phi\,.
\end{align*}
In order to have $\left(D_{\mu}\phi\right)^{\prime}=\gamma D_{\mu}\phi$,
$f$ and $g$ must obey 
\[
f=\frac{1}{\alpha}g\,,\;\partial_{\mu}f=g\partial_{\mu}\varphi\,.
\]
The solutions are either $f=\alpha^{-1}e^{\alpha\varphi}$ and $g=e^{\alpha\varphi}$
or $f=e^{\alpha\varphi}$ and $g=\alpha e^{\alpha\varphi}$. The second
pair is favored, since it provides the correct vanishing torsion limit
for the charge current density, as we explain below. Therefore, the
scalar fields $\phi$ must transform as 
\[
\phi^{\prime}=\exp\left(iqe^{\alpha\varphi}\epsilon\right)\phi
\]
and the covariant derivative is given by 
\[
D_{\mu}\phi=\left(\partial_{\mu}-iq\alpha e^{\alpha\varphi}A_{\mu}\right)\phi\,.
\]
An action functional invariant by (\ref{eq:local-gauge-transf}) has
the form: 
\[
S_{\phi}=\int d^{4}x\sqrt{-g}\,\overline{D_{\mu}\phi}D^{\mu}\phi\,.
\]
The equations of motion are 
\[
\frac{1}{\sqrt{-g}}D_{\mu}\left(\sqrt{-g}D^{\mu}\phi\right)=0\,.
\]
By Noether's theorem one can write the current density 
\[
\epsilon j_{\phi}^{\mu}=\frac{\partial\mathcal{L}}{\partial\partial_{\mu}\phi}\delta_{\epsilon}\phi+\frac{\partial\mathcal{L}}{\partial\partial_{\mu}\overline{\phi}}\delta_{\epsilon}\overline{\phi}\,,
\]
where $\delta_{\epsilon}\phi=iqe^{\alpha\varphi}\epsilon\phi$ and
$\delta_{\epsilon}\overline{\phi}=-iqe^{\alpha\varphi}\epsilon\overline{\phi}$.
Therefore, the gauge invariant current is 
\begin{align}
j_{\phi}^{\mu} & =iq\sqrt{-g}e^{\alpha\varphi}\phi\overline{D^{\mu}\phi}-iq\sqrt{-g}e^{\alpha\varphi}\overline{\phi}D^{\mu}\phi=-iq\sqrt{-g}e^{\alpha\varphi}\overline{\phi}\overleftrightarrow{D^{\mu}}\phi\nonumber \\
 & =-i\sqrt{-g}qe^{\alpha\varphi}\left(\overline{\phi}\partial^{\mu}\phi-\partial^{\mu}\overline{\phi}\phi\right)-2\sqrt{-g}\alpha q^{2}e^{2\alpha\varphi}A^{\mu}\overline{\phi}\phi\,.\label{current}
\end{align}
It is easily seen that 
\[
\left(\partial_{\mu}-\alpha T_{\mu}\right)j_{\phi}^{\mu}=-iqe^{\alpha\varphi}\partial_{\mu}\left(\sqrt{-g}\,\overline{\phi}\,\overleftrightarrow{D^{\mu}}\phi\right)\,.
\]
From the equations of motion for $\phi$ and $\overline{\phi}$, one
can show that $\partial_{\mu}\left(\sqrt{-g}\,\overline{\phi}\,\overleftrightarrow{D^{\mu}}\phi\right)=0$.
Thus, onshell, one has the generalized continuity equation, see equation
(\ref{eq:conserv-corrente}) in the Appendix.

\subsection{Physical fields and the action\label{sub:Physical-fields-and}}

We also note that the action (\ref{eq:A2}) has symmetry (\ref{eq:HRRS-gauge}),
which, with the notation introduced in this section, reads 
\[
\delta_{\epsilon}^{\prime}A_{\mu}=e^{-\alpha\varphi}\partial_{\mu}\epsilon\,,\,\,\,\delta_{\epsilon}^{\prime}F_{\mu\nu}=0\,.
\]
This symmetry, proposed in \cite{Hojman1978,Hojman1979} is actually
the usual $U\left(1\right)$-symmetry for the redefined vector potential
\begin{equation}
B_{\mu}=e^{\alpha\varphi}A_{\mu}\,,\label{eq:HRRSb-field}
\end{equation}
since $\delta_{\epsilon}^{\prime}B_{\mu}=\partial_{\mu}\epsilon$
and $F_{\mu\nu}=e^{-\alpha\varphi}H_{\mu\nu}$, where $H_{\mu\nu}=\partial_{\mu}B_{\nu}-\partial_{\nu}B_{\mu}$.

For the proposed new gauge symmetry, one can similarly redefine $A_{\mu}$,
so as to obtain a $U\left(1\right)$-like symmetry instead of (\ref{eq:local-gauge-transf}).
In this case, one has 
\begin{equation}
b_{\mu}=\alpha e^{\alpha\varphi}A_{\mu}\,,\,\,\delta_{\epsilon}b_{\mu}=\partial_{\mu}\left(e^{\alpha\varphi}\epsilon\right)\,.\label{eq:b-field}
\end{equation}
It is clear this is not the usual $U\left(1\right)$-symmetry, since
the torsion scalar $\varphi$ participates in the transformation.
Nonetheless, we can also write $h_{\mu\nu}=\alpha e^{\alpha\varphi}F_{\mu\nu}$,
where $h_{\mu\nu}=\partial_{\mu}b_{\nu}-\partial_{\nu}b_{\mu}$ is
the usual Maxwell field tensor.

The current (\ref{current}) has clearer meaning in terms of the $b$-field
(\ref{eq:b-field}), since then $j_{\phi}^{\mu}$ becomes 
\[
j_{\phi}^{\mu}=\sqrt{-g}e^{\alpha\varphi}\left[-iq\left(\overline{\phi}\partial^{\mu}\phi-\partial^{\mu}\overline{\phi}\phi\right)-2q^{2}b^{\mu}\overline{\phi}\phi\right]\,.
\]
In the above form the geometrical contributions are clearly separated,
and one attains the interpretation of a field $\phi$ of charge $q$
coupled to the Maxwell connection $b_{\mu}$. In the limit of vanishing
torsion, $\varphi\rightarrow0$, the current becomes the usual charged
Klein-Gordon field minimally coupled to $b_{\mu}$, which is invariant
under the usual $U\left(1\right)$ gauge transformation. Unlike in
the HRRS model, the current density is affected by torsion even in
the absence of electromagnetic fields.

From the above considerations, we take the physical electromagnetic
four-vector to be $b_{\mu}$ and the gauge-invariant physical electromagnetic
field strength tensor to be $h_{\mu\nu}$. This will have important
consequences, as we show in the next section on the Newtonian limit.

Apart from a total divergence, the Einstein action is 
\[
S_{g}=\frac{1}{16\pi}\int d^{4}x\sqrt{-g}\left(\bar{R}-6\alpha^{2}T^{2}\right)\,,
\]
where $\bar{R}$ is the curvature scalar defined in terms of the Levi-Civita
connection $\bar{\nabla}$, and we have definition (\ref{eq:newdef})
for the trace part of the torsion tensor. Therefore, the total action
encompassing gravitation, electromagnetism and matter fields is given
by the integral of the Lagrange density 
\begin{equation}
\mathcal{L}=\frac{\sqrt{-g}}{16\pi}\left(\bar{R}-6\alpha^{2}\partial_{\mu}\varphi\partial^{\mu}\varphi-\frac{e^{-2\alpha\varphi}}{\alpha^{2}}h^{2}-4\overline{D_{\mu}\phi}D^{\mu}\phi\right)\,.\label{eq:lagrange-density}
\end{equation}
We note that the kinetic term for the torsion scalar field has the
right sign, according to \cite{Popawski2006}. The Euler-Lagrange
equations following (\ref{eq:lagrange-density}) are 
\begin{align}
\delta b_{\mu} & :\left(\bar{\nabla}_{\mu}-\alpha\partial_{\mu}\varphi\right)\frac{e^{-\alpha\varphi}}{\alpha}h^{\mu\nu}+iq\alpha e^{\alpha\varphi}\left(\overline{D^{\nu}\phi}\phi-\overline{\phi}D^{\nu}\phi\right)=0\,,\label{41}\\
\delta\varphi & :3\alpha^{2}\bar{\square}\varphi+\frac{1}{\alpha}\bar{\nabla}_{\mu}\left(e^{-2\alpha\varphi}h^{\mu\nu}b_{\nu}\right)-i\alpha qb^{\mu}\left(\overline{\phi}D_{\mu}\phi-\overline{D_{\mu}\phi}\phi\right)=0\,,\label{42}\\
\delta\phi & :\bar{\nabla}^{\mu}\overline{D_{\mu}\phi}+iq\overline{D^{\mu}\phi}b_{\mu}=0\,,\label{43}\\
\delta\overline{\phi} & :\bar{\nabla}^{\mu}D_{\mu}\phi-iqb_{\mu}D^{\mu}\phi=0\,.\label{44}
\end{align}
Using equation (\ref{41}) in the equation (\ref{42})\ for the torsionic
scalar $\varphi$ gives 
\[
3\alpha\bar{\square}\varphi+\frac{1}{2}\frac{e^{-2\alpha\varphi}}{\alpha^{2}}h_{\mu\nu}h^{\mu\nu}=0\,.
\]
We note that taking $\varphi=0$, all torsion-dependent terms in the
Lagrangian density (\ref{eq:lagrange-density}) vanish and one recovers
the usual equations for electromagnetism and matter fields in general
relativity in terms of $f_{\mu\nu}$.

For $\alpha=-1$ the above expressions are identical to the ones derived
from the HRRS model. Therefore, at first sight, it seems that our
theory is equivalent to the HRRS theory by the transformation 
\[
\alpha\varphi\rightarrow-\varphi~,\ \alpha q\rightarrow q\,.
\]
Even the new gauge transformation (\ref{eq:local-gauge-transf}) can
be made $\alpha$-independent by means of the rescaling $\frac{\epsilon}{\alpha}\rightarrow\epsilon$,
\[
\delta_{\epsilon}A_{\mu}=\partial_{\mu}\epsilon-\partial_{\mu}\varphi\epsilon\,.
\]
However, we must stress two points: First, we take (\ref{eq:local-gauge-transf}) 
as the gauge transformation defining measurable quantities $b_{\mu}$ and 
$h_{\mu\nu}$. As a result,
measurable fields and gauge-invariant fields in our approach differ
from those in the HRRS (\ref{eq:HRRSb-field}). The second point is
that making the above rescalings in order to eliminate $\alpha$ is
tantamout to setting it to $-1$, which is not acceptable according
to available experimental data. Thus, this constant needs to be determined
by experiment. We show in the next section that the value $\alpha=-1$
can be excluded and the present theory for $\alpha\neq-1$ is nonequivalent
to the HRSS one.

\subsection{Newtonian Limit and comparison with experimental results}

In this section we quote results from \cite{Ni1979} with regard to
test body accelerations in the Newtonian limit of the HRRS model.
From a merely formal viewpoint, their results can be translated into
our model by means of the map $\varphi\mapsto-\alpha\varphi$ and
the definition of the (measurable) field $b_{\mu}$ from (\ref{eq:b-field})
and its associated strength tensor $h_{\mu\nu}=\partial_{\mu}b_{\nu}-\partial_{\nu}b_{\mu}$.
By solving the equations of motion for $\varphi$ in the weak field
approximation, as well as considering the background electric and
magnetic fields of the sun, one arrives at a relation between $\varphi$
and the sun's gravitational potential $U$. On the other hand, the
test body acceleration in a local inertial frame can be computed from
the Lagrangian density (\ref{eq:lagrange-density}) as 
\[
\ddot{X}_{\mu}=-\frac{2}{\alpha}\frac{E_{e}-E_{m}}{m}\partial_{\mu}\varphi\,,
\]
where $m$, $E_{m}$ and $E_{e}$ are the mass, the total electric
and magnetic energy of the test body. Combining the two results, and
using typical values of $E_{e}$ for platinum and aluminum atoms,
one gets the deviation in the acceleration of platinum ($\ddot{X}_{i}^{\mathrm{Pt}}$)
and aluminum ($\ddot{X}_{i}^{\mathrm{Al}}$) atoms: 
\[
\ddot{X}_{i}^{\mathrm{Pt}}-\ddot{X}_{i}^{\mathrm{Al}}=\alpha^{-4}2\times10^{-7}\partial_{i}U\,.
\]

The above deviation vanishes with a precision of $1$ part in $10^{12}$
of $\partial_{i}U$ according to \cite{Nobili2013}. It is clear that
the HRRS model is in disagreement with experimental data, since it
corresponds to $\alpha=-1$. In the case of our theory, the above
experimental result can be used to establish a lower bound for $\alpha$,
\[
\left(2\times10^{-7}\right)\alpha^{-4}<10^{-12}\Rightarrow\left\vert \alpha\right\vert \gtrsim20\,.
\]

It is important to note that making the rescalings indicated at the
end of section (\ref{sub:Physical-fields-and}) and proceeding as
above to obtain the Newtonian limit, the constant $\alpha$ disappears.
This is a consequence of the choice of the physical fields (\ref{eq:b-field}).

\section{Final remarks and perspectives}

We have proposed a gauge-invariant model of propagating torsion which
couples to the Maxwell field and to charged particles. Our model requires
the introduction of a constant $\alpha$ into the definition of the
trace part of the torsion tensor, so that minimal coupling can be
achieved at the level of the Action. We provide in the Appendix a
realization of the equations of motion in terms of the semi-minimal
coupling (coupling at the level of differential forms), in which case
the Maxwell equations are formally identical to the torsionless Maxwell
equations by the substitution of the exterior derivative $d$ by an
appropriate map $D$.

The fact that we do not have MCP when applied to the equations of
motion does not represent a setback, since, as was pointed out in
\cite{Hehl2001}, MCP can only be safely applied at the Action
level, because of the possible appearance of curvature-dependent terms
in the equations of motion, violating the equivalence principle. We
can expect that, as in the case of the classical limit of quantum
theories, different theories with torsion correspond to the same flat
space limit. This is a different viewpoint from the one adopted in
\cite{Mosna2005}, where MCP is restricted to map equivalent theories
in flat space to equivalent theories in curved space.

Despite the formal identification between the HRRS model and our present
proposal, which can be achieved by mapping the torsion scalar $\varphi$
to $-\alpha\varphi$ and the charge $q$ to $\alpha q$, in our construction
$\alpha$ naturally appears as coupling constant in the construction
of the gauge-invariant Maxwell-like action (\ref{eq:A2}), in terms
of the new gauge transformation (\ref{eq:local-gauge-transf}). In
this sense, the HRRS model can be seen as a particular case when $\alpha=-1$.
However, current experimental tests rule out the value $\alpha=-1$,
and set a lower bound on $\alpha$, $\left\vert \alpha\right\vert >20$.

Besides, in the case of HRRS theory, one can see that the proposed
gauge transformation (\ref{eq:HRRS-gauge}) can be recast in familiar
terms by means of a field redefinition, $B_{\mu}=e^{-\phi}A_{\mu}\Rightarrow\delta_{\epsilon}B_{\mu}=\partial_{\mu}\epsilon$;
while the gauge transformation proposed here is nontrivial in the
sense that the gauge parameter involves the torsion field, $b_{\mu}=\alpha e^{\alpha\varphi}A_{\mu}\Rightarrow\delta_{\epsilon}b_{\mu}=\partial_{\mu}\left(e^{\alpha\varphi}\epsilon\right)$.
We also note that the $\alpha^{-1}$ term in the gauge transformation
(\ref{eq:local-gauge-transf}) is not essential, since, as is the
case in usual Maxwell theory, one can redefine the gauge field $A_{\mu}$
such that the gauge transformation becomes $\delta_{\epsilon}A_{\mu}=\partial_{\mu}\epsilon+\alpha\partial_{\mu}\varphi$.
Then, as a result, the Maxwell action will be multiplied by a factor
of $\alpha^{-2}$. Perhaps in this sense $\alpha$ might be construed
to be some sort of charge carried by torsion, which is related to
the interaction between the torsion field and other fields in the
model. In fact, by setting $\alpha$ to zero, one negates any and
all torsionic effects.

Finally, we wish to note that, albeit the introduction of the constant
$\alpha$ in the definition of the gauge transformations might appear
as a trivial modification in both HRRS and Saa's theory, this modification,
without the introduction of the new gauge transformation, is not enough
to allow both theories to dodge the available experimental restrictions.
In addition, in our case the introduction of $\alpha$ is a matter
of principle, that led to a nontrivial coupling between torsion and
electromagnetism and without which the MCP is ill-defined (see Section
\ref{sec:Minimal-coupling}).

A natural sequel to this work would be to apply the theory here presented
to a specific cosmological model, in order to obtain testable physical
predictions and hopefully an upper bound for $\alpha$, and also to
generalize the gauge principle to non-Abelian gauge groups.

\paragraph*{Acknowledgements:}

We would like to thank Alberto Saa, Jose Abdalla Helayël-Neto and
Ilya Shapiro for insightful remarks and useful comments. TSP thanks
Conselho Nacional de Desenvolvimento Científico e Tecnológico (CNPq)
for financial support.

\appendix

\section{Semi-minimal coupling\label{sub:Semi-minimal-coupling}}

In the following we consider the general gauge transformation and
modified Maxwell field tensor given by (\ref{eq:local-gauge-transf})
and (\ref{eq:field-tensor}), respectively. For arbitrary $\alpha$,
we show that a minimal coupling on the level of differential forms
can be achieved which circumvents issues related to restrictions on
the connection coefficients and the failure of MCP when applied to
the homogeneous equations encountered in \cite{Saa1993}.

The Maxwell field tensor (\ref{eq:field-tensor}) can be written invariantly
in terms of the differential two-form 
\[
F=dA-\alpha T\wedge A\,,
\]
where $A=A_{\mu}dx^{\mu}$ and $T=T_{\mu}dx^{\mu}$ are the vector
potential and torsion trace one-forms. Consider the map $D:\Lambda^{p}\rightarrow\Lambda^{p+1}$
defined in the space of $p$-forms $\omega$ such that%
\footnote{We note that $D$ is not a graded derivation, i.e., one does not have
$D\left(\omega_{p}\wedge\omega_{q}\right)=D\omega_{p}\wedge\omega_{q}+\left(-1\right)^{p}\omega_{p}\wedge D\omega_{q}$,
where $\omega_{p}$ is a $p$-form and $\omega_{q}$ is a $q$-form.%
} 
\[
D\omega=d\omega-\alpha T\wedge\omega\,.
\]
One can show that for an arbitrary $p$-form $\omega$ 
\[
D^{2}\omega=0
\]
since $T$ is exact, $T=d\varphi$. From the nilpotency of the map
$D$ it follows that 
\[
DF\equiv0\,,
\]
which is the analog of the homogeneous Maxwell equations $df\equiv0$,
where $f=dA$. The homogeneous equations in a local coordinate map
are 
\[
\left(\partial_{\mu}-\alpha T_{\mu}\right)F_{\nu\sigma}+\left(\partial_{\sigma}-\alpha T_{\sigma}\right)F_{\mu\nu}+\left(\partial_{\nu}-\alpha T_{\nu}\right)F_{\sigma\mu}\equiv0\,.
\]
Thus, the homogeneous equations are identically zero, and no restriction
on either $F$ or $T$ arise.

The analog of the inhomogeneous Maxwell equations without sources,
$\ast d\ast f=0$ is 
\[
\ast D\ast F=\left(-\bar{\nabla}_{\mu}F^{\mu\nu}+\alpha T_{\mu}F^{\mu\nu}\right)dx_{\nu}=0\,,
\]
which coincides with the equations of motion (\ref{eq:equations-motion}).

We have thus shown that the semi-minimal coupling given by the substitution
of the de-Rahm exterior product $d$ by the map $D$ provides the
Maxwell equations coupled to the trace part of the torsion tensor:
\begin{gather*}
f=dA\rightarrow F=DA\,,\\
df\equiv0\rightarrow DF\equiv0\,,\\
\ast d\ast f=0\rightarrow\ast D\ast F=0\,.
\end{gather*}

\section{Maxwell Equations with sources}

In this section we apply the formalism presented in the previous section
in order to calculate conserved currents.

Let us introduce the current density $3$-form 
\[
j=\frac{1}{3!}j^{\mu}\varepsilon_{\mu\nu\kappa\sigma}dx^{\nu}\wedge dx^{\kappa}\wedge dx^{\sigma}\,,
\]
such that the inhomogeneous Maxwell equations become 
\begin{equation}
D\ast F=j\,.\label{eq:maxwell-with-sources}
\end{equation}
Since $D^{2}=0$, one has the condition $Dj=0$ to ensure consistency
of the Maxwell equations, which in a local chart reads 
\begin{equation}
\left(\bar{\nabla}_{\mu}-\alpha T_{\mu}\right)j^{\mu}=0\,.\label{eq:conserv-corrente}
\end{equation}
The interaction term in the action is gauge invariant provided the
current satisfies the conservation equation (\ref{eq:conserv-corrente}):
\[
\delta_{\epsilon}\int d^{4}x\sqrt{-g}j^{\mu}A_{\mu}=-\frac{1}{\alpha}\int d^{4}x\sqrt{-g}\left(\bar{\nabla}_{\mu}-\alpha T_{\mu}\right)j^{\mu}\,.
\]

Now consider the three-form $\tau_{T}=*i_{T}\ast F$, where $i_{T}$
is the interior derivative along the vector field $T$, which in coordinates
is given by 
\[
\tau_{T}=\frac{\alpha}{3!}\left(T_{\mu}f_{\nu\sigma}+T_{\sigma}f_{\mu\nu}+T_{\nu}f_{\sigma\mu}\right)\, dx^{\mu}\wedge dx^{\nu}\wedge dx^{\sigma}\,.
\]
This three-form is covariantly conserved: 
\[
D\tau_{T}=d\tau_{T}-\alpha T\wedge\tau_{T}\equiv0\,,
\]
which is consistent with $DF\equiv0$. Since $T\wedge\tau_{T}$ vanishes,
it follows that $\tau_{T}$ is a closed form, 
\begin{equation}
d\tau_{T}=0\,.\label{eq:closed-3form}
\end{equation}
Thus one can construct a conserved quantity, the gauge invariant one-form
$j_{T}=\ast\tau_{T}$, which has the local expression 
\[
j_{T\mu}=-\frac{\alpha}{2}\varepsilon_{\mu\nu\rho\kappa}T^{\nu}F^{\rho\kappa}\equiv-\frac{\alpha}{2}\varepsilon_{\mu\nu\rho\kappa}T^{\nu}f^{\rho\kappa}\,.
\]
Following (\ref{eq:closed-3form}), one has $d\ast j_{T}=0$: 
\[
\partial_{\mu}j_{T}^{\mu}\equiv\nabla_{\mu}j_{T}^{\mu}\equiv0\,.
\]
Thus, if $\Sigma$ is space-like hypersurface, one has the conserved
quantity 
\[
Q=\alpha\int_{\Sigma}d^{3}x\sqrt{\tilde{g}}\mathbf{T}\cdot\mathbf{B}\,,
\]
where $\tilde{g}_{\mu\nu}$ is the induced metric on $\Sigma$, and
$T$ and $B$ are torsion and magnetic field vectors. One can show
that the total charge $Q$ is a boundary term and vanishes at infinity
in case the fields have vanishing boundary values at infinity, as
in the case of propagating torsion theory. From $j_{T}$ one can construct
a dual torsionic source for electromagnetism, called ``electric current\textquotedblright \ in
\cite{Hojman1978}, $j_{E}=i_{T}F$, which in coordinates reads $j_{E}^{\mu}=-\alpha F^{\mu\nu}T_{\nu}$.

 \bibliographystyle{unsrt}
\bibliography{library,\string"/Users/fresneda/work/mendeley.bib/library\string"}

\end{document}